\documentclass[runningheads]{llncs}
\usepackage{amsmath}
\usepackage{graphicx}
\begin{document}
\title{Non-Orthogonal HARQ-CC over SDR: A GNU Radio-Based Implementation\thanks{This work was supported in part by National Natural Science Foundation of China under Grants 62171200,  62171201, and 92367201, in part by Guangdong Basic and Applied Basic Research Foundation under Grant 2023A1515010900, in part by Fundamental Research Funds for the Central Universities under Grant 21625361, in part by the Science and Technology Development Fund, Macau SAR under Grant 0020/2025/RIB1. (Corresponding Authors: Jintao Wang, Zheng Shi.)}}
%
%
\author{Hongling Huang\inst{1}
\and
Jintao Wang\inst{1} \and
Zheng Shi\inst{1} 
\and Xu Wang\inst{1}
\and Guanghua Yang\inst{1}
\and Shaodan Ma\inst{2}
\and Haichuan Ding \inst{3}}
\authorrunning{H. Hongling, W. Jintao et al.}
%
\institute{School of Intelligent Systems Science and Engineering, Jinan University, Zhuhai, China \and
State Key Laboratory of Internet of Things for Smart City, University of Macau, Macau, China \and
School of Cyberspace Science and Technology, Beijing Institute of Technology, Beijing, China\\
Email: \email{curryhhl@stu.jnu.edu.cn}, \email{wang.jintao@connect.um.edu.mo}, \email{zhengshi@jnu.edu.cn}, \email{xuwang@jnu.edu.cn}, \email{ghyang@jnu.edu.cn}, \email{shaodanma@um.edu.mo}, \email{hcding@bit.edu.cn} }

\maketitle              
\begin{abstract}
Hybrid Automatic Repeat Request (HARQ) schemes typically allocate all available resources to retransmit failed packets to ensure reliability. However, under stringent delay constraints, these schemes often exhibit low spectral efficiency and increased transmission latency. To address these challenges, this paper proposes an efficient Non-Orthogonal HARQ with Chase Combining (N-HARQ-CC) transmission strategy. Specifically, the proposed approach allocates a larger portion of retransmission resources to new data packets, reserving only a small fraction for retransmitting previously erroneous packets. This is based on the observation that only a small number of information bits are typically incorrect, enabling surplus communication resources to be utilized for transmitting new messages. The N-HARQ-CC scheme retransmits the same redundant version of a failed packet and employs Maximum Ratio Combining (MRC) for decoding. To minimize complex packet scheduling and decoding complexity, the proposed scheme limits superposition to at most two messages per transmission round. At the receiver, Successive Interference Cancellation (SIC) is used to decouple the superimposed messages. The proposed N-HARQ-CC system was implemented using GNU Radio and USRP platforms for validation. Compared to conventional Type-I HARQ and HARQ-CC schemes, the proposed scheme achieves a significant improvement in spectral efficiency of approximately 0.5 bps/Hz, aligning with the low-latency requirements of 6G networks.

\keywords{GNU Radio  \and hybrid automatic repeat request \and non-orthogonal multiple access \and software-defined radio \and USRP.}
\end{abstract}
\section{Introduction}
Compared with the fifth-generation (5G) mobile communication system, which primarily focuses on three typical scenarios—enhanced mobile broadband (eMBB), ultra-reliable low-latency communication (URLLC), and massive machine-type communication (mMTC)—the sixth-generation (6G) mobile communication technology not only inherits the core capabilities of 5G but also introduces several key enhancements. Specifically, 6G incorporates further enhanced mobile broadband (FeMBB) to support terabit-per-second (Tbps) data rates; extremely reliable low-latency communication (ERLLC) to meet the stringent demands of critical applications; and ultra-massive machine-type communication (umMTC) to achieve significantly higher global device connectivity \cite{b1}.

As 6G imposes more stringent requirements on latency and reliability, conventional hybrid automatic repeat request (HARQ) mechanisms are expected to face unprecedented challenges. HARQ, a key reliability enhancement mechanism in 5G, improves robustness by combining forward error correction and retransmission strategies. However, in the context of 6G URLLC scenarios, conventional HARQ schemes suffer from excessive feedback and queuing delays, limiting their suitability for latency-critical applications such as industrial automation, vehicular communications, and holographic services.

To address these limitations, various advanced HARQ schemes have been proposed. For multicast channels and delay-sensitive services, fixed-rate HARQ schemes based on superposition coding (SC), as well as multiplexed HARQ (M-HARQ), were shown to enhance throughput and spectral efficiency via power allocation and iterative interference cancellation \cite{b2,b3}. Furthermore, multi-packet HARQ (MP-HARQ) was proposed to allow multiple packets to share the same radio resource. Through dynamic coding strategies guided by Markov decision processes (MDPs), MP-HARQ achieves significant performance gains in high spectral efficiency scenarios \cite{b4,b5}.

On the other hand, hybrid non-orthogonal multiple access (NOMA) systems enable energy-efficient user scheduling through joint optimization of power allocation and block length \cite{b6}. Grant-free NOMA (GF-NOMA) and power-domain NOMA (PD-NOMA) are also considered as key enablers for 6G \cite{b7,b8}. These approaches facilitate the integration of multi-packet transmission with HARQ schemes, improving both system capacity and reliability \cite{b9,b10}. 

Non-orthogonal HARQ (N-HARQ) mechanism enables simultaneous transmission of new and retransmitted packets within the same time slot via power-domain superposition. This significantly reduces queuing delays, thereby enhancing delay performance \cite{b11,b12}. Moreover, dynamic adjustment of the maximum number of retransmissions allows a flexible trade-off between latency and reliability, resulting in improved overall system throughput.
In terms of practical implementation, numerous downlink NOMA systems have been experimentally validated on software-defined radio (SDR) platforms using GNU Radio and universal software radio peripheral (USRP)  \cite{b13,b14,b15}. However, most HARQ and N-HARQ schemes remain in the theoretical or simulation stage with limited real-world deployment or system-level validation, highlighting the need for further experimental research.

In this work, we propose a practical  N-HARQ with Chase
Combining (N-HARQ-CC) scheme that integrates non-orthogonal SC with HARQ mechanisms. The N-HARQ-CC scheme retransmits the same redundant version of a failed packet and employs Maximum Ratio Combining (MRC) for decoding. By superimposing user packets in the power domain, the proposed scheme achieves enhanced throughput while supporting low-latency retransmissions. Furthermore, we implement and validate the proposed mechanism using a combination of GNU Radio and USRP hardware, demonstrating its practical feasibility and performance advantages.

\section{System Model}
This section introduces the proposed N-HARQ-CC scheme first, with which the procedure of signal reconstruction and the analysis of signal-to-interference-plus-noise (SINR) are presented.

\subsection{N-HARQ-CC Scheme}
This paper investigates an N-HARQ-aided downlink communication system with one transmitter and one receiver. Unlike traditional HARQ schemes, which retransmit only the redundancy information of a failed message, the N-HARQ scheme incorporates a new information message during retransmissions. This approach leverages the fact that only a small portion of the information bits are typically erroneous, allowing surplus communication resources to be utilized for transmitting the next message. Consequently, the N-HARQ scheme enhances spectral efficiency and reduces transmission delay, albeit with a slight increase in computational complexity. In this paper, we propose a practical N-HARQ scheme, termed as N-HARQ-CC, which capitalizes on both chase combining and superposition coding. In particular, similar to the HARQ-CC scheme, our proposed scheme assumes that the same redundant information of failed messages is retransmitted and MRC is used at the receiver. Moreover, to avoid a complex packet scheduling mechanism and high decoding complexity, at most two messages are allowed to be superimposed at each transmission round. At the receiver, the technique of successive interference cancellation (SIC) is applied to separate the two superimposed messages. Moreover, to avoid network congestion, the maximum number of HARQ rounds for each message is limited to $M$. According to whether a positive or negative acknowledgement (ACK/NACK) message is fed back to the transmitter, the following two transmission modes are triggered by the proposed N-HARQ-CC scheme.
\subsubsection{Initial Transmission Mode}
If there is no NACK feedback or the maximum number of HARQ rounds for each message is reached, a new HARQ cycle begins. Accordingly, a new message will be delivered in the first HARQ round, i.e., initial transmission mode (ITM). The received packet in the initial HARQ round can be written as
\begin{equation}
\mathbf{y}_{1}=h_1\sqrt{P}\mathbf{x}_1+{\bf n}_{1},
\end{equation}
where ${\bf x}_1$ refers to the encoded message with normalized average power, $P$ denotes the transmission power, $h_1$ refers to the channel coefficient, and $n_l$ represents the complex additive white Gaussian noise (AWGN) with zero mean and unity variance. The decoding is performed at the receiver according to the observation $\mathbf{y}_{1}$. If the receiver successfully recovers $\mathbf{x}_1$, an ACK message will be sent back and a new HARQ cycle will be started. Otherwise, if an NACK is returned, a retransmission mode will be triggered subsequently, that is, a new message will be superimposed with the failed message in the power domain in a retransmission HARQ round. 
\subsubsection{Retransmission Mode}
A retransmission mode (RM) is initiated if at least one NACK message is received, unless the maximum number of HARQ rounds for each message has been reached. In the retransmission mode, two messages are superimposed at the transmitter. Without loss of generality, we denote by the two messages $\mathbf{x}_{k_1}$ and $\mathbf{x}_{k_2}$ with $k_1 < k_2$. For expositional simplicity, we refer to $\mathbf{x}_{k_1}$ and $\mathbf{x}_{k_2}$ as the old and new packets, respectively. Hence, the received signal in the $l$-th HARQ round can be written as 
\begin{equation}\label{eqn:lsign}
\mathbf{y}_{l}=h_l\left(\alpha\sqrt{ P}\mathbf{x}_{k_1}+\sqrt{(1-\alpha^2)P}\mathbf{x}_{k_2}\right)+\mathbf{n}_{l},
\end{equation}
where $h_l$ corresponds to the channel coefficient, $\alpha$ represents the power allocation coefficient of the old packet, and ${\bf n}_l$ denotes the complex AWGN. Since there is a small amount of erroneous information bits, more power can be allocated to the new packet. According to the principle of NOMA, the new packet $\mathbf{x}_{k_2}$ is decoded first. After successfully reconstructing  $\mathbf{x}_{k_2}$, the SIC is applied to decode the old packet $\mathbf{x}_{k_1}$ by subtracting $\mathbf{x}_{k_2}$ from $\mathbf{y}_{l}$. If the receiver successfully decodes both messages, a new cycle will start in the next HARQ round. Otherwise, the retransmission mode continues and the delivered messages (old message $\mathbf{x}_{\rm old}$ and new message $\mathbf{x}_{\rm new}$) in the next HARQ round are given by
\begin{equation}
[\mathbf{x}_{\rm old},\mathbf{x}_{\rm new}]=
    \begin{cases}
    \left[\mathbf{x}_{k_1},\mathbf{x}_{k_2}\right],  & \mathcal F_O^{\rm NACK} \wedge \mathcal F_N^{\rm NACK}\\
      \left[\mathbf{x}_{k_2},\mathbf{x}_{k_2+1}\right],  &\mathcal F_O^{\rm ACK} \wedge \mathcal F_N^{\rm NACK}\\
      \left[\mathbf{x}_{k_1},\mathbf{x}_{k_2+1}\right],  &\mathcal F_O^{\rm NACK} \wedge \mathcal F_N^{\rm ACK} 
    \end{cases},
\end{equation}
where $\mathcal F_O^{\rm ACK} $ and $\mathcal F_O^{\rm NACK} $ denote the success and the failure of recovering the old message, respetively, and the similar definitions are also applicable to $\mathcal F_N^{\rm ACK} $ and $\mathcal F_N^{\rm NACK} $ for the new message.

According to the transmission mechanism of the N-HARQ-CC scheme, the previously received packets and the currently received packet can be jointly decoded through MRC. Furthermore, the new message is decoded first due to its higher power allocation coefficient and the old message is then decoded by leveraging the operation of SIC. Hence, the signal reconstruction and SINR analysis for these two messages are discussed separately.
\subsection{Signal Reconstruction and SINR Analysis}
\subsubsection{Decoding of the New Message}
To proceed, we assume that the new message has been transmitted for $m_2$ HARQ rounds. In the $l$-th HARQ round, the receiver attempts to recover the new message ${\bf x}_{k_2}$ by combining the received signals from the latest $m_2$ HARQ rounds, i.e., $\mathbf{y}_{l-m_2+1},\cdots,\mathbf{y}_{l}$. By employing MRC, the combined signal is written as 
\begin{equation}\label{eqn:mrc_new}
    \mathbf{y}_{m_2}^{\rm MRC}=\frac{1}{\left\| {{{\bf{h}}_{l - m_2 + 1:l}}} \right\|^2}\sum_{i=l-m_2+1}^{l}{h_i^*}\mathbf{y}_{i},
\end{equation}
where ${{{\bf{h}}_{l -m_2 + 1:l}}}=(h_{l-m_2+1},\cdots,h_{l})^{\rm T}$. 
Substituting \eqref{eqn:lsign} into \eqref{eqn:mrc_new} leads to 
\begin{align}\label{eqn:mrc_n}
    \mathbf{y}_{m_2}^{\rm MRC} =& \sqrt {(1 - {\alpha ^2})P} {{\bf{x}}_{{k_2}}} + \alpha \sqrt P {{\bf{x}}_{{k_1}}} \notag\\
    &+ \frac{1}{{{{\left\| {{{\bf{h}}_{l - m_2 + 1:l}}} \right\|}^2}}}\sum\limits_{i = l - m_2 + 1}^l {h_i^*{{\bf{n}}_i}} ,
\end{align}
Since the old message is directly treated as noise, the received SINR for the new message can be derived as 
\begin{equation}
    \gamma_{m_2}^{\rm new} = \frac{{(1 - {\alpha ^2})P{{\left\| {{{\bf{h}}_{l - m_2 + 1:l}}} \right\|}^2}}}{{\alpha^2 P{{\left\| {{{\bf{h}}_{l - m_2 + 1:l}}} \right\|}^2} + 1}}.\label{(10)}
\end{equation}

\subsubsection{Decoding of the Old Message}
After recovering the new message ${\bf x}_{k_2}$, the old message ${\bf x}_{k_1}$ is decoded by using SIC. For notational simplicity, we assume that the old message ${\bf x}_{k_1}$ has been delivered for $m_1$ times. Since messages that fail after the maximum number of transmissions cannot be canceled, they will interfere with the decoding of the old message \(\mathbf{x}_{k_1}\). Clearly, at most one such failure event presents due to the restriction on the number of HARQ rounds, i.e., \(M\). Depending on whether a single failed SIC operation occurs, we categorize the subsequent discussion into two cases, i.e., no SIC failure and one SIC failure.
\paragraph{No SIC Failure}
This case indicates that all the messages from ${\bf x}_{k_1+1}$ to ${\bf x}_{k_2}$ can be cancelled out. Hence, subtracting the interfered signals ${\bf x}_{k_1+1},\cdots,{\bf x}_{k_2}$ from ${\bf y}_{l-m_1+1},\cdots,{\bf y}_{l}$ yields $\tilde {\bf y}_{l-m_1+1},\cdots,\tilde  {\bf y}_{l}$ that only contain the desired signal \(\mathbf{x}_{k_1}\).
Therefore, $\tilde {\bf y}_{i}$ can be written as
\begin{equation}
    \tilde {\bf y}_{i} =h_i \alpha\sqrt{ P} \mathbf{x}_{k_1} + \mathbf{n}_{i}, \, i\in [l-m_1+1,l].
\end{equation}
By applying MRC for $\tilde {\bf y}_{l-m_1+1},\cdots,\tilde  {\bf y}_{l}$, the combined signal can be written as
\begin{equation}\label{eqn:mrc_old}
    \tilde{\mathbf{y}}_{m_1}^{\rm MRC}=\frac{1}{\left\| {{{\bf{h}}_{l - m_1 + 1:l}}} \right\|^2}\sum_{i=l-m_1+1}^{l}{h_i^*}\tilde{\mathbf{y}}_{i}.
\end{equation}
Accordingly, the received SNR for decoding the old message \(\mathbf{x}_{k_1}\) can be calculated as
\begin{equation}\label{eqn:mrc_oldsnr}
       \gamma_{m_1}^{\rm old} = \alpha^2 P{\left\| {{{\bf{h}}_{l - m_1 + 1:l}}} \right\|^2}.
\end{equation}

\begin{figure}[t]
  \centering
  \includegraphics[width=1\textwidth, keepaspectratio]{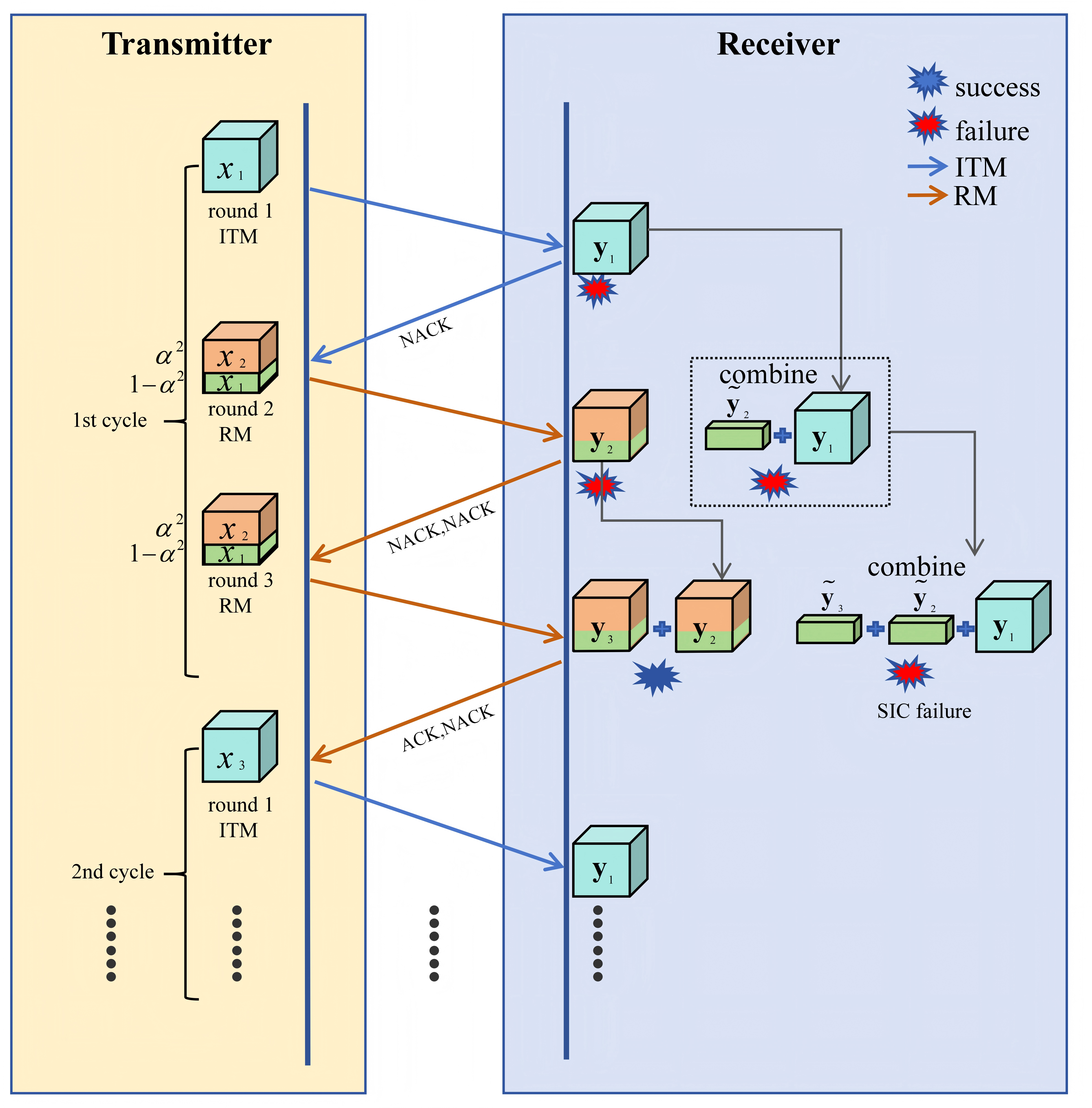} 
  \caption{The example of one SIC failure in decoding the old message with $M=3$.}
  \label{HARQ_flow}
\end{figure}

\paragraph{One SIC Failure}
When an SIC failure occurs for a given message, the transmission round count of the other message in the superimposed signal is necessarily smaller, since the new packet is only superimposed on the previous message that requires retransmission, and there exists only one such message at a time. Therefore, it can be concluded that during the entire HARQ process for a given message, at most one message may experience SIC failure. For simplicity, we assume that the SIC failure occurs at the $l_1$-th HARQ round, and the message ${\bf x}_{k_0}$ is discarded after $M$ HARQ rounds. Clearly, all the messages from ${\bf x}_{k_1+1}$ to ${\bf x}_{k_2}$ have been successfully recovered and can be subtracted from ${\bf y}_{l_1+1}$ to ${\bf y}_l$. Whereas, the message ${\bf x}_{k_0}$ cannnot be cancelled out from ${\bf y}_{l-m_1+1}$ to ${\bf y}_{l_1}$ and interferes the recovery of the old signal ${\bf x}_{k_1}$. Accordingly, the combined signal can be written as
\begin{align}\label{eqn:mrc_old2}
    \tilde {\bf{y}}_{{m_1}}^{{\rm{MRC}}} =& \frac{1}{{{{\left\| {{{\bf{h}}_{l - {m_1} + 1:l}}} \right\|}^2}}}\left( {\sum\limits_{i = l - {m_1} + 1}^{{l_1}} {h_i^*{{\bf{y}}_i}}  + \sum\limits_{i = {l_1} + 1}^l {h_i^*{{\tilde {\bf{y}}}_i}} } \right)  \notag\\
    =&\alpha \sqrt P {{\bf{x}}_{{k_1}}} + \frac{{{{\left\| {{{\bf{h}}_{l - {m_1} + 1:{l_1}}}} \right\|}^2}}}{{{{\left\| {{{\bf{h}}_{l - {m_1} + 1:l}}} \right\|}^2}}}\sqrt {(1 - {\alpha ^2})P} {{\bf{x}}_{{k_2}}} \notag\\
    &+ \frac{1}{{{{\left\| {{{\bf{h}}_{l - {m_1} + 1:l}}} \right\|}^2}}}\sum\limits_{i = l - {m_1} + 1}^l {h_i^*{{\bf{n}}_i}} .
\end{align}
As a consequence, the received SINR for the old message is given by 
\begin{equation}
    \gamma _{{m_1}}^{{\rm{old}}} = \frac{{\alpha^2 P{{\left\| {{{\bf{h}}_{l - {m_1} + 1:l}}} \right\|}^4}}}{{(1 - {\alpha ^2})P{{\left\| {{{\bf{h}}_{l - {m_1} + 1:{l_1}}}} \right\|}^4} + {{\left\| {{{\bf{h}}_{l - {m_1} + 1:l}}} \right\|}^2}}}.
\end{equation}
The illustration of one SIC failure in decoding the old message with $M=3$ is presented in Fig. \ref{HARQ_flow}.

\section{SDR-Based Implementation of N-HARQ-CC}\label{IISP}
\begin{figure}[t]
  \centering
  \includegraphics[width=1\textwidth, keepaspectratio]{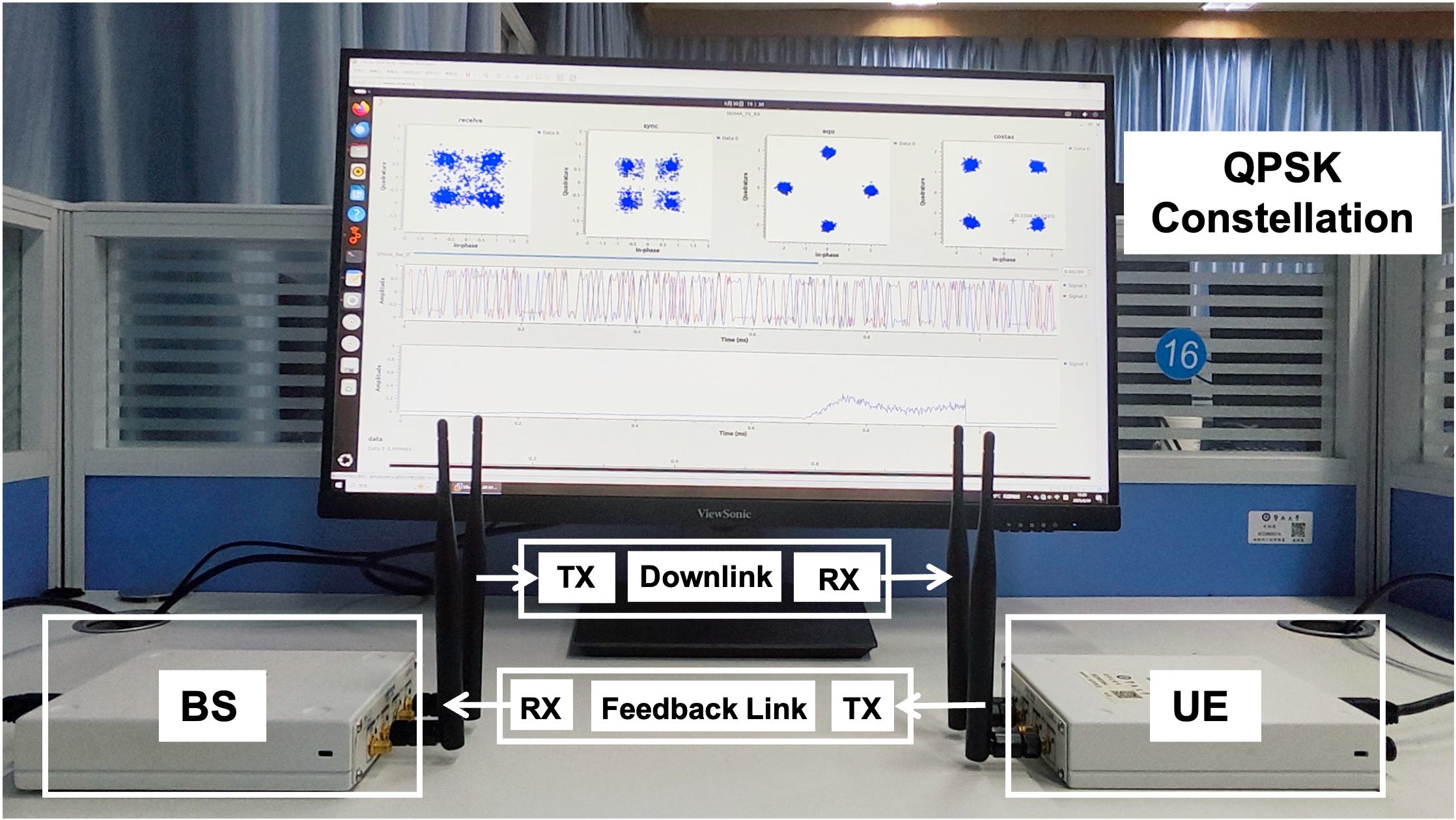} 
  \caption{Experimental setup of N-HARQ-CC transmission over SDR using two USRP B210 devices.}
  \label{USRP}
\end{figure}

\begin{figure}[htbp]
  \centering
  \includegraphics[width=1\textwidth, keepaspectratio]{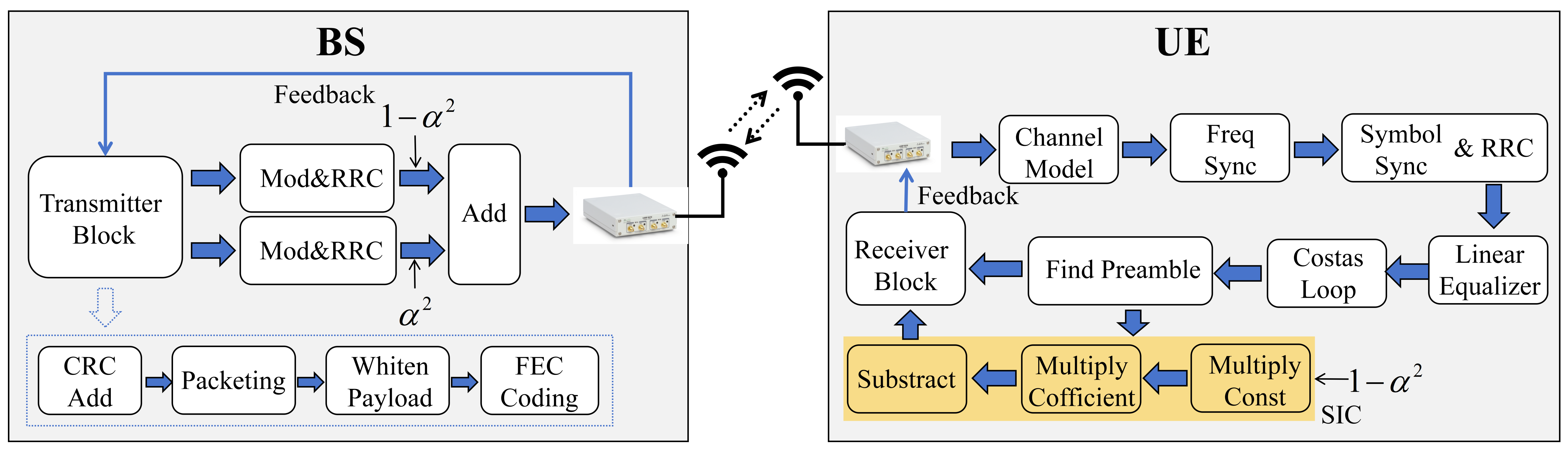} 
  \caption{System architecture of the proposed N-HARQ-CC scheme based on GNU Radio.}
  \label{SDR}
\end{figure}

\begin{figure*}
  \centering
\includegraphics[width=1\textwidth, keepaspectratio]{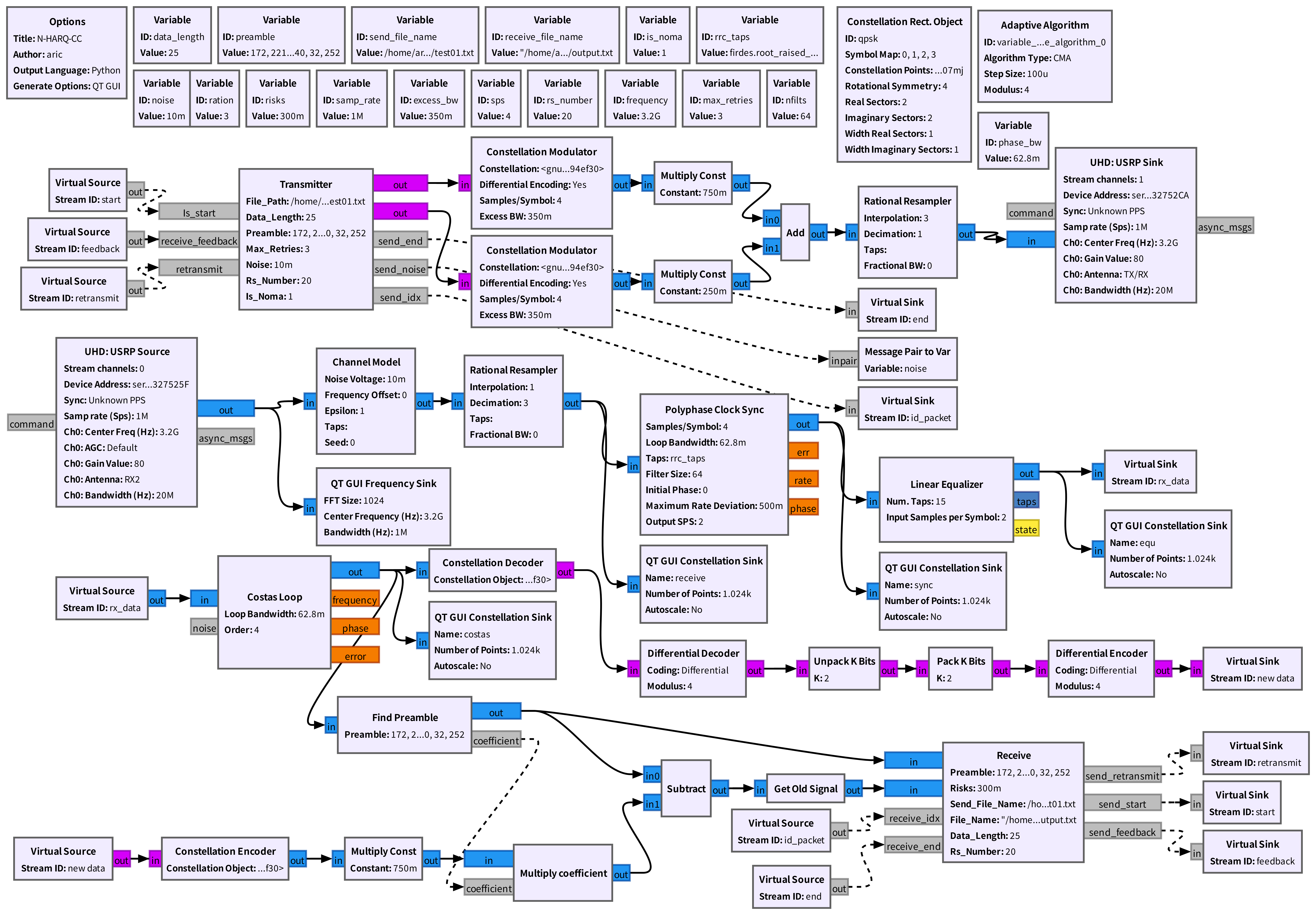} 
  \caption{GNU Radio flowgraph of the proposed N-HARQ-CC through 2 USRPs.}
  \label{flowgraph}
\end{figure*}
GNU Radio is an open-source software development toolkit that offers a wide range of signal processing modules. It enables flexible construction and implementation of SDR systems. When integrated with hardware platforms such as the USRP, GNU Radio serves as an ideal tool for rapid prototyping, experimental validation, and cognitive radio research\cite{b16}.

In this study, we develop a SC transmission framework to support two data packets, which is implemented on the GNU Radio platform using two USRP B210 devices. As illustrated in Fig.~\ref{USRP}, one USRP B210 serves as the BS and the other as the UE. Both USRPs are configured with two channels to construct the downlink transmission and uplink feedback paths.
The system architecture of the proposed N-HARQ-CC scheme is shown in Fig.~\ref{SDR} and the GNU Radio flowgraph of the transmitter and receiver is presented in Fig. \ref{flowgraph}.

\subsection{Transmitter}
 On the transmitter side, the core module is a custom Python Embedded Block, referred to as the \textit{Transmitter Block}, which is responsible for data packet construction and automatic repeat request (ARQ) control. The input file is first segmented into multiple packets with fixed byte lengths. Each packet is appended with a cyclic redundancy check (CRC) checksum for error detection, followed by the insertion of fixed-format headers and tails to complete the packetization. To enhance the randomness and ensure uniform bit distribution, whitening is applied to the CRC, header, and tail bits. The whitened packet is then encoded using forward error correction (FEC) to improve error resilience.

In the ARQ control module, packets are selectively transmitted based on feedback results and predefined retransmission strategies. The selected packets are modulated into quadrature phase shift keying (QPSK) symbols and processed with a root-raised cosine (RRC) filter. 
The resulting superimposed signal is transmitted via the USRP B210 to the receiver. And the CRC code has a length of 32 bits, the FEC employs a Reed–Solomon (RS) code with a coding rate of approximately 0.6, and the roll-off factor of the RRC filter is set to 0.35.

\subsection{Receiver}
On the receiver side, the wireless signal is captured using the USRP B210. To facilitate performance evaluation under various SNR conditions, a Channel Model block is used to introduce AWGN. The received signal then undergoes a series of preprocessing steps, including frequency synchronization, symbol synchronization, and RRC filtering. To mitigate channel distortion, a linear equalizer based on the Constant Modulus Algorithm (CMA) is employed for channel equalization. Subsequently, a Costas loop is applied to achieve carrier phase synchronization, thereby improving demodulation accuracy. Finally, SIC is performed to separate the superimposed signals.
\begin{figure}[t]
  \centering
  \begin{minipage}[b]{0.7\textwidth}
    \centering
    \includegraphics[width=\linewidth, keepaspectratio]{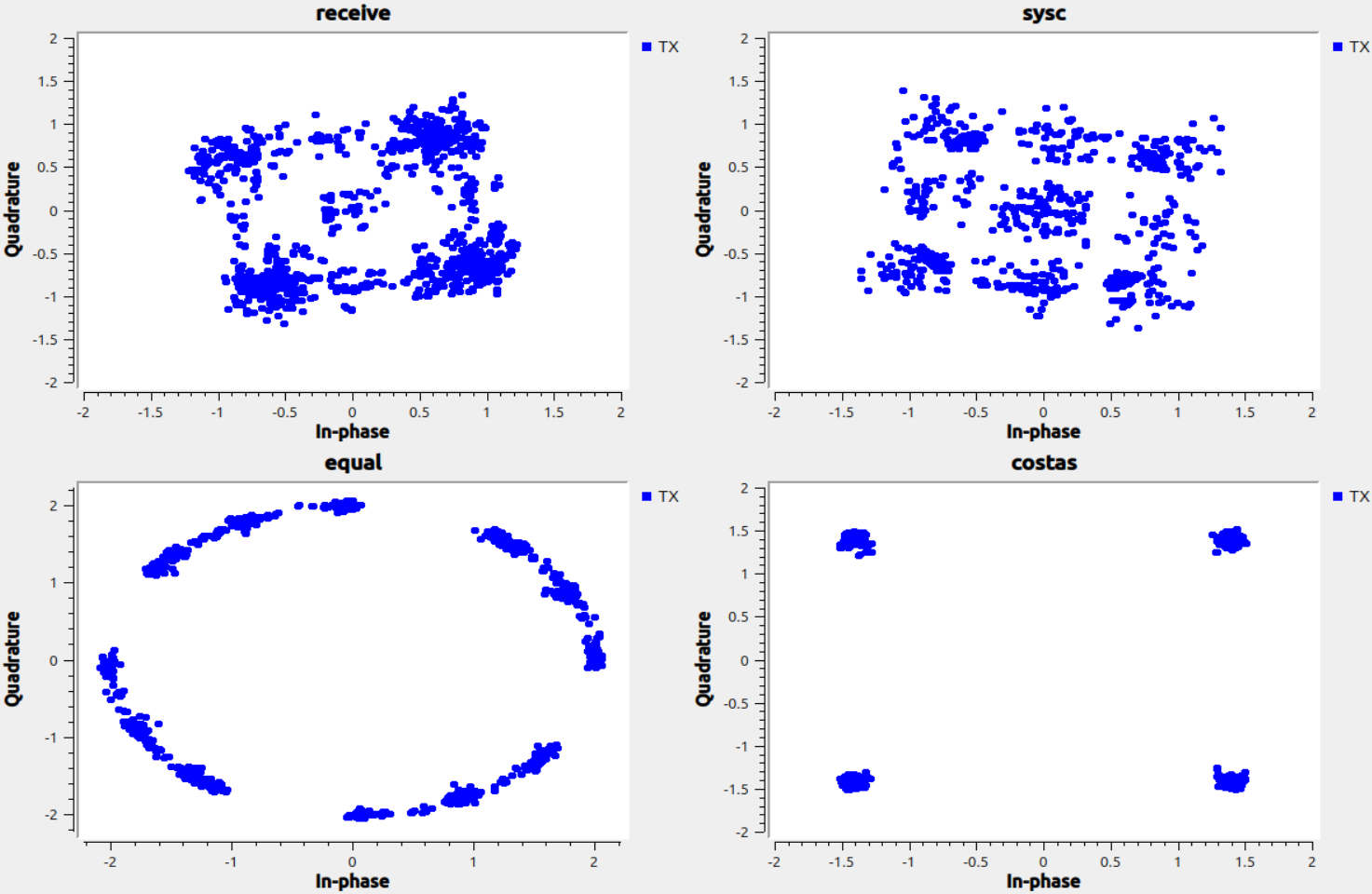}
    \caption{QPSK constellation in ITM mode.}
    \label{qpsk1}
  \end{minipage}
  \hfill
  \begin{minipage}[b]{0.7\textwidth}
    \centering
    \includegraphics[width=\linewidth, keepaspectratio]{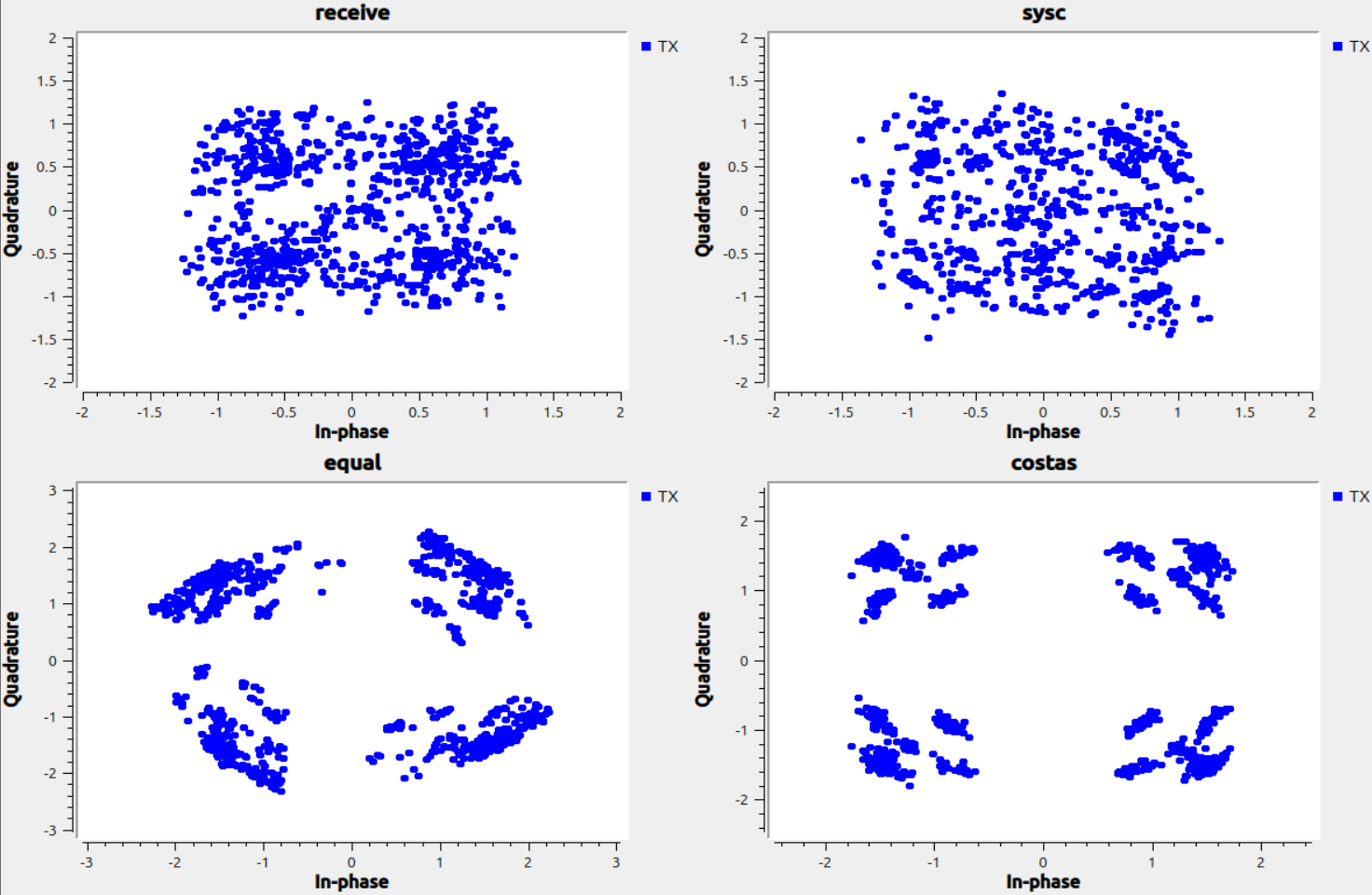}
    \caption{QPSK constellation in RM mode.}
    \label{qpsk2}
  \end{minipage}
\end{figure}

The demodulated signals are passed to the custom \textit{Receiver Block}, which serves as the core processing module at the receiver. It is responsible for data packet parsing and HARQ feedback generation. During packet parsing, the signal is first decoded using FEC to recover the original bit stream, followed by dewhitening. The packet header is then checked for integrity. If the header is invalid, the packet is discarded; otherwise, a CRC check is performed to verify data integrity. Based on the CRC result, the system determines whether retransmission is required and generates the corresponding HARQ feedback message.

\section{Experimental Results}\label{ERWHU}
In this experiment, the SDR platform is employed for both transmission and reception to evaluate system performance under varying channel conditions. 
To ensure stable operation within a band-limited channel, the experimental setup is configured with a 3.2 GHz carrier frequency, 1 MHz sampling rate, and 20 MHz channel bandwidth.
Each transmitted frame contains 200 information bits. A total of 1757 packets are transmitted to ensure statistically meaningful results for performance evaluation. 
To investigate the impact of different SNR levels on system performance, AWGN is introduced through the channel block to enable controlled variation of the SNR. For each SNR level, the BER and the normalized average throughput per frame are measured, providing a comprehensive performance profile of the system across different channel qualities.

Fig.~\ref{qpsk1} and Fig.~\ref{qpsk2} depict the QPSK constellation diagrams for the ITM and RM transmission modes, respectively. Each figure presents four constellation diagrams that capture the signal at successive stages of the receiver processing chain: (i) the received signal at the user, (ii) the signal after synchronization and RRC filtering, (iii) the output after CMA equalization, and (iv) the result after phase and frequency correction using the Costas loop.

\begin{figure}[t]
  \centering
  \includegraphics[width=0.8\textwidth, keepaspectratio]{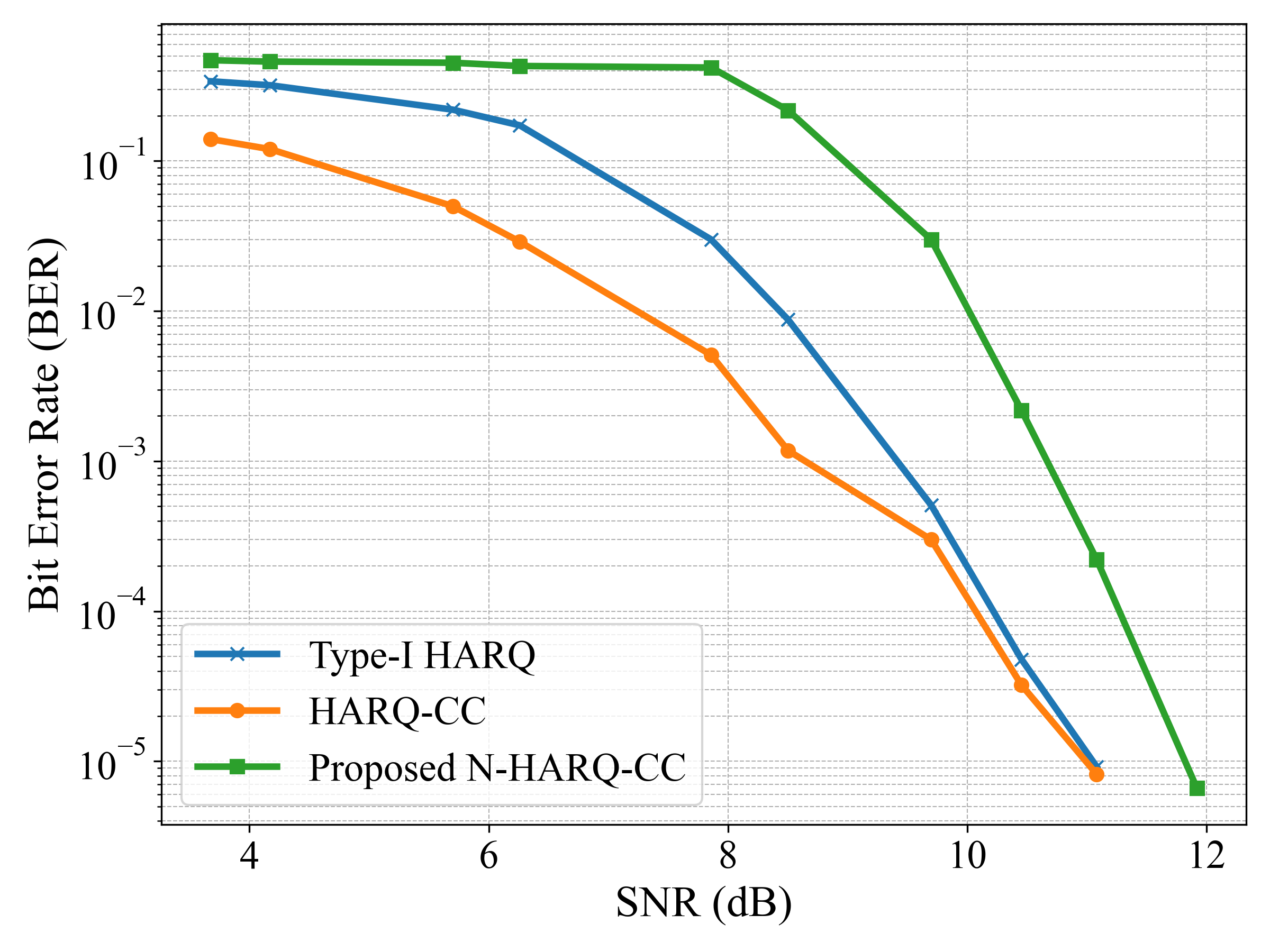} 
  \caption{BER analysis of conventional and proposed transmission schemes.}
  \label{ber}
\end{figure}

 Fig.~\ref{ber} presents the BER performance of three transmission schemes under varying SNR conditions. Generally, the BER of all schemes decreases as the SNR increases, which is consistent with the theoretical expectation.
Among the compared schemes, the Type-I HARQ exhibits significantly inferior BER performance relative to the HARQ-CC scheme in the low-to-moderate SNR range (approximately 4–10 dB). This is attributed to the fact that Type-I HARQ only performs simple retransmissions upon initial decoding failure, thereby failing to exploit the decoding gain provided by soft combining. In contrast, the MRC achieves better BER performance by leveraging soft combining of multiple received copies. As the SNR increases, the BER performance of both schemes converges in the high-SNR regime, ultimately approaching zero. This indicates that under favorable channel conditions, the relative benefits of retransmission strategies and coding gains diminish, as error-free transmission becomes more readily achievable.

In comparison, the proposed scheme experiences notable BER degradation in the low-to-moderate SNR region. This is primarily attributed to inter-packet interference and resource allocation dilution inherent in parallel packet transmission, which effectively reduces the received SNR. A significant improvement in BER is observed only when the SNR exceeds approximately 10 dB. This behavior indicates that while the RM can enhance system throughput.

\begin{figure}[t]
  \centering
  \includegraphics[width=0.8\textwidth, keepaspectratio]{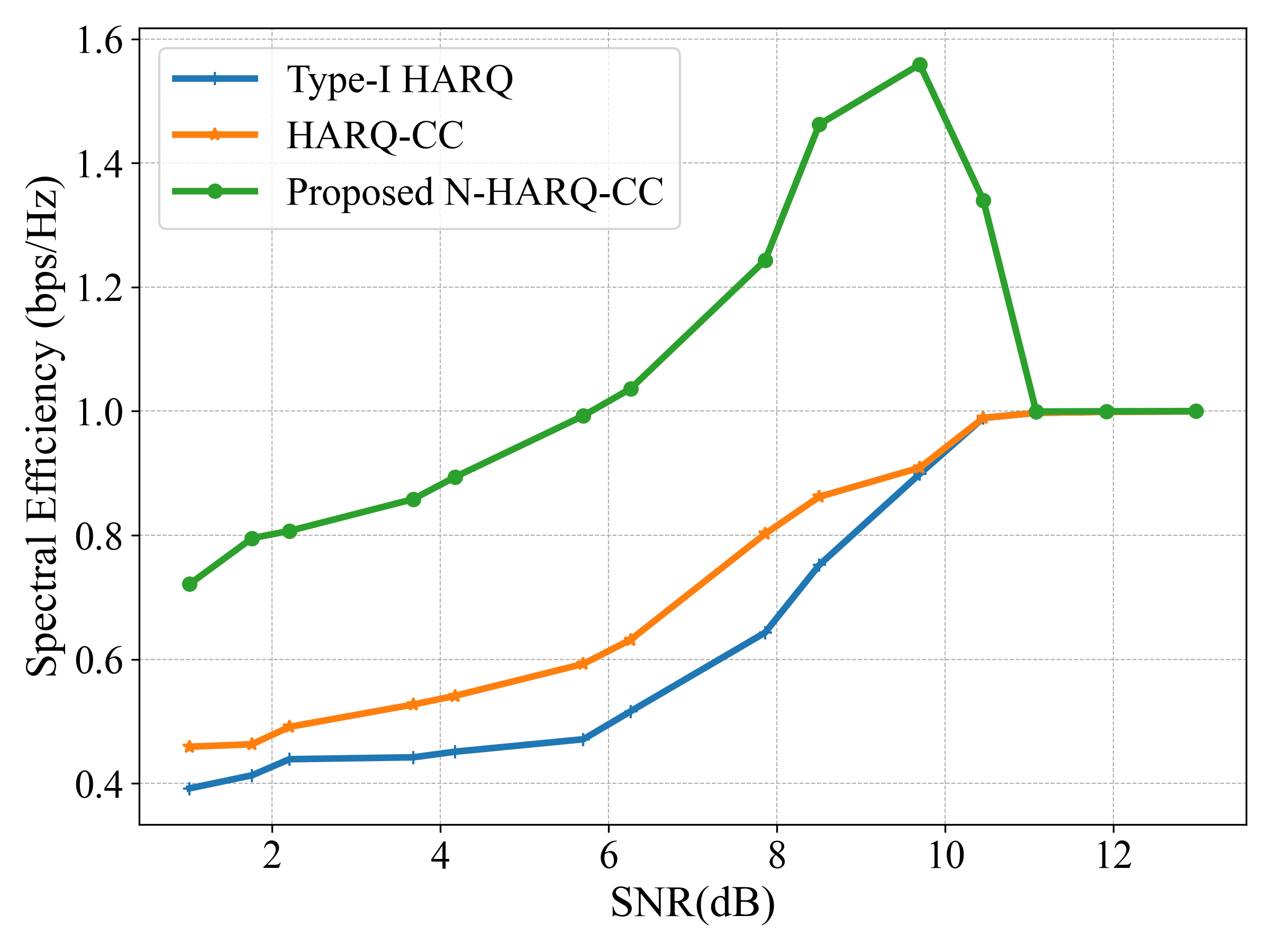} 
  \caption{Spectral efficiency analysis of conventional and proposed transmission schemes.}
  \label{throughput}
\end{figure}

Fig.~\ref{throughput} illustrates the spectral efficiency (SE) of three representative transmission schemes under varying SNR conditions. This metric reflects how efficiently data bits are transmitted relative to the theoretical maximum.
At low SNR levels, all three schemes exhibit poor SE performance due to high error rates. As the SNR increases, the SE of both the HARQ-CC and Type-I HARQ schemes gradually improves, saturating around 11 dB and approaching the theoretical limit. This saturation indicates that most packets are successfully decoded on the first attempt. However, the SE of Type-I HARQ remains slightly lower than that of HARQ-CC, as it does not exploit soft combining gains. The proposed N-HARQ-CC scheme reaches its peak SE at moderate SNRs (around 10 dB), but the SE then declines and stabilizes at higher SNRs. This behavior is attributed to that improved channel conditions reduce the retransmission frequency, thereby causing the system to revert to RM and thus diminishing the throughput benefit of parallel ITM operation.
Overall, the proposed N-HARQ-CC scheme improves spectral efficiency by around 0.5 bps/Hz over the conventional HARQ-CC scheme. Its dynamic adaptability makes it particularly suitable for low latency in 6G networks.

\section{CONCLUSION}\label{IISP}
This paper proposed an efficient N-HARQ-CC transmission strategy to address the limitations of traditional HARQ schemes, namely, low spectral efficiency and increased transmission latency under stringent delay constraints. The N-HARQ-CC scheme retransmitted the same redundant version of a failed packet with a superimposed new packet and the SIC technique was used to decouple the superimposed messages. Furthermore, we implemented the proposed N-HARQ-CC system using GNU Radio and USRP platforms for validation. The experimental results demonstrated that the proposed N-HARQ-CC achieves significantly higher throughput compared to Type-I HARQ and HARQ-CC schemes. 

Furthermore, this study reveals a robustness challenge under poor channel conditions, where fixed power allocation can lead to the SIC failure and sustained multi-packet superposition, thereby limiting throughput. This observation underscores the necessity for joint optimization of dynamic power allocation and adaptive transmission mode switching (e.g., fallback to ITM), potentially leveraging reinforcement learning frameworks, which is a primary direction for our subsequent investigation.

%
%
%
%

\end{document}